\documentclass{notices}

\usepackage{amssymb}
\usepackage{amsmath}
\usepackage{bm}
\usepackage{xcolor}
\usepackage{url}
\usepackage{algorithm}
\usepackage{algpseudocode}
\usepackage{graphicx}
\usepackage{stmaryrd}
\SetSymbolFont{stmry}{bold}{U}{stmry}{m}{n}

\title{
Mathematical modeling of the vascular system
}

\author{
  Ju Liu
  \affil{
		Ju Liu is an assistant professor in the Department of Mechanics and Aerospace Engineering at Southern University of Science and Technology. His email address is liuj36@sustech.edu.cn.
    }
  \and
  Ingrid S. Lan
  \affil{
    Ingrid S. Lan is a graduate student in the Department of Bioengineering at Stanford University. Her email address is ingridl@stanford.edu.
   }
  \and
  Alison L. Marsden
  \affil{
    Alison L. Marsden is a professor and Vera Moulton Wall Center faculty scholar in the Departments of Pediatrics (Cardiology), Bioengineering, and Institute for Computational \& Mathematical Engineering at Stanford University. Her email address is amarsden@stanford.edu.
   }
}

\date{}

\begin{document}

\maketitle

\subsection*{The vascular network}
The cardiovascular system is a closed circuit in which the heart pumps blood from the left and right ventricles to the systemic and pulmonary circulations. The two circulations work synchronously to deliver oxygenated blood to the body and the heart itself and to route deoxygenated blood to the lungs for reoxygenation. In each circulation, blood travels away from the heart through the larger elastic arteries and smaller muscular arteries and arterioles before reaching the capillary network; within the capillaries, an exchange of oxygen, carbon dioxide, nutrients, and metabolites occurs between the blood and surrounding tissue; finally, blood returns to the heart through venules and veins. The distensibility of the vessels gives rise to the propagation of pressure and flow waves through the fluid at a finite speed. Importantly, since the diameters of capillaries and red blood cells are of the same order, the multiphase nature of blood must be considered when modeling the microcirculation. In this article, we restrict our discussion to mathematical modeling of the larger vessels with a continuum approach. Interested readers may refer to \cite{quarteroni2019mathematical} for a discussion on mathematical modeling of the heart.

% physiological background here
The vascular network exhibits complex topological, geometric, physiologic, and material properties that differ greatly from subject to subject, necessitating a patient-specific approach to mathematical modeling. Furthermore, blood vessels constitute an intricate biological system that dynamically adapts to the body's changing needs. In conjunction with the nervous system, the vascular endothelium \textit{acutely} regulates the vessel tone and thus the distribution of flow to different tissues. In response to sustained changes in hemodynamic loading, arteries \textit{chronically} adapt in the form of growth and remodeling (G\&R), yielding changes in both the geometry and material properties. Experimental studies have, for example, demonstrated the tendency of the arterial wall to thicken in response to sustained increases in blood pressure and that of the lumen to enlarge in response to sustained increases in flow \cite{Humphrey2002}. 

\begin{figure*}
	\begin{center}
	\begin{tabular}{c}
\includegraphics[angle=0, trim=0 0 0 0, clip=true, scale = 0.28]{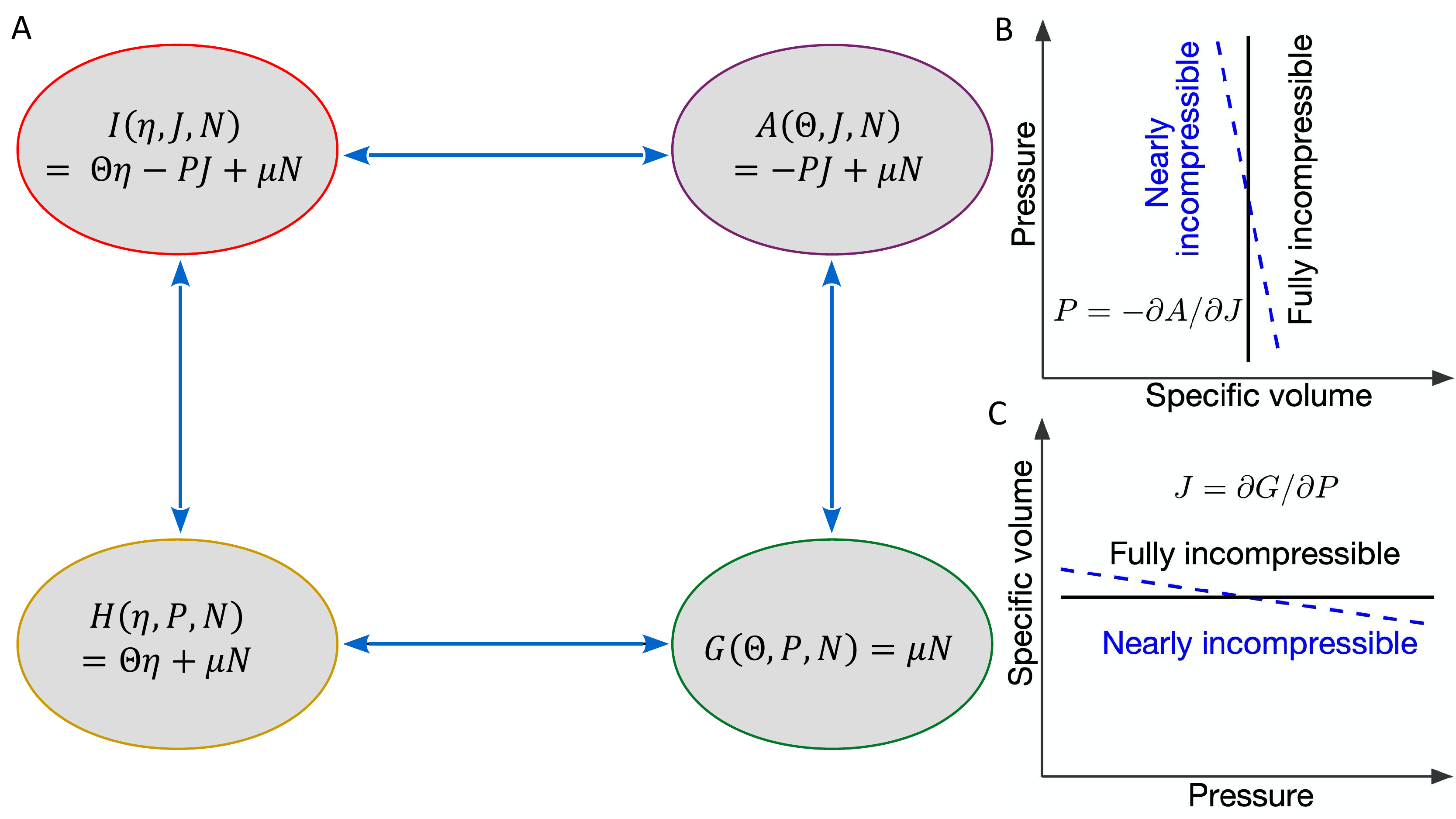}
\end{tabular}
\end{center}
\caption{ (A) Illustration of thermodynamic potentials: the internal energy $I$, Helmholtz free energy $A$, Gibbs free energy $G$, and enthlapy $H$. The blue arrows indicate their relations via Legendre transformations. (B) The infinite slope of the pressure-volume curve illustrates the inability of the Helmholtz energy to describe incompressible behavior. (C) The Helmholtz free energy can be transformed into the Gibbs free energy via a Legendre transformation. The resulting system has a saddle-point nature, which requires special mathematical techniques for analysis.} 
\label{fig:thermodynamics}
\end{figure*}

% disease, diagnosis, and treatment
Proper functioning of the cardiovascular system can, however, be disrupted by either congenital or acquired diseases. These range from structural abnormalities present at birth (e.g. underdeveloped ventricles, blockages, and holes) on the congenital front, to structural changes that develop (e.g.  atherosclerotic stenoses and aneurysms) on the acquired front. Despite profound advances in treatments and medical imaging, cardiovascular diseases remain the leading cause of death worldwide, and congenital heart defects remain a leading cause of birth defect-associated infant illness and death in the United States. Gold-standard interventions are still met with troubling complications. In particular, stents are frequently met with restenosis, bypass grafts experience alarmingly high rates of failure, and many diseases require a ``watch and wait" treatment strategy. As alternative interventional approaches are explored, there is a pressing need to quantitatively understand and predict disease initiation and progression as well as failure modes observed in the clinic. The performance of novel medical device designs and surgical approaches must also be accurately predicted and evaluated.

It is clear that hemodynamics, vascular wall biomechanics, and cellular biochemical responses are all deeply intertwined, and comprehensive mathematical modeling of the vascular network on all scales presents tremendous research opportunities. Just as modern-day aeronautical and automotive design rely heavily on predictive mathematical modeling, we envision a future in which patient-specific modeling of cardiovascular diseases becomes a routine component of preventive care, diagnostic care, and treatment planning. In this article, we discuss recent research efforts and our perspectives on the most pressing open research directions in mathematical modeling of the vascular system.

\subsection*{A unified continuum modeling framework}
Cardiovascular modeling requires consideration of multiphysics phenomena, including the interaction of fluids, solids, and other relevant physics, such as electrophysiology, active contraction, or the transport of reactive chemical species. Computational modeling of fluid and solid sub-problems has conventionally employed dichotomous formulations for the two sub-domains, creating challenges and disparities for modeling the coupled system. Unifying the formulations in a single continuum framework would indeed be ideal. Traditionally, the constitutive relations for finite elasticity have been described in terms of the Helmholtz free energy (also known as the strain energy), likely because it is a function of quantities that can be conveniently measured and controlled in laboratories, including temperature, specific volume, the number of molecules. The Helmholtz free energy, however, ceases to be a proper potential in the incompressible limit, as the specific volume becomes constrained as a constant; accordingly, ill-conditioned matrix problems arise in numerical analyses for incompressible materials. To address this issue, a Legendre transformation can be performed on the Helmholtz free energy with respect to the specific volume (Figure \ref{fig:thermodynamics}). The resulting thermodynamic potential, namely the Gibbs free energy, can then be used to describe material behavior in both the incompressible and compressible regimes. For incompressible flows, it yields the incompressible Navier-Stokes equations; for compressible flows, the Gibbs free energy leads to the pressure primitive variable formulation, which has been deemed robust for all-speed flows. Following classical principles of mechanics, the governing equations based on the Gibbs free energy can be written in the arbitrary Lagrangian-Eulerian (ALE) description as follows, 
\begin{align}
\label{eq:isothermal_governing_eqn_mass}
& 0 = \beta_{\theta} \left. \frac{\partial p}{\partial t} \right|_{\bm \chi} + \beta_{\theta} \left(\bm v - \hat{\bm v} \right)\cdot \nabla_{\bm x} p + \nabla_{\bm x} \cdot \bm v, \displaybreak[2] \\
\label{eq:isothermal_governing_eqn_momentum}
& \bm 0 = \left. \rho \frac{\partial \bm v}{\partial t} \right|_{\bm \chi} + \rho \left(\bm v - \hat{\bm v} \right) \cdot \nabla_{\bm x} \bm v - \nabla_{\bm x} \cdot \bm \sigma_{\mathrm{dev}} + \nabla_{\bm x} p - \rho \bm b,
\end{align}
wherein $\rho$ is the density, $p$ is the pressure, $\beta_{\theta} := (\partial \rho / \partial p) / \rho$ is the isothermal compressibility coefficient, $\bm \chi$ represents the spatial coordinate with respect to a suitable reference frame that is chosen case by case and that moves with domain velocity $\hat{\bm v}$, $\bm x$ is the spatial coordinate in the Eulerian domain, $\bm v$ is the velocity, $\bm \sigma_{\textup{dev}}$ is the deviatoric component of the Cauchy stress, and $\bm b$ is the body force per unit mass. Whereas the volumetric behavior is completely described by $\rho(p)$ and $\beta_{\theta}(p)$, the isochoric behavior is determined by $\bm \sigma_{\textup{dev}}$. \textit{Constitutive modeling} refers to the design of an appropriate form for the Gibbs free energy, which dictates the analytical forms of $\rho(p)$, $\beta_{\theta}(p)$, and $\bm \sigma_{\textup{dev}}$ and therefore closes the system. As will be revealed, equations \eqref{eq:isothermal_governing_eqn_mass}-\eqref{eq:isothermal_governing_eqn_momentum} properly describe both viscous fluids and elastic solids, and we are actively working to extend the framework to inelastic solids as well. This unified framework bridges the gap between techniques for computational fluid dynamics (CFD) and structural dynamics, and further simplifies the analysis of fluid-structure interaction (FSI) coupled problems by presenting a unified problem requiring only a single solution strategy \cite{Liu2018}.

\begin{figure*}
	\begin{center}
	\begin{tabular}{c}
\includegraphics[angle=0, trim=20 290 0 0, clip=true, scale = 0.4]{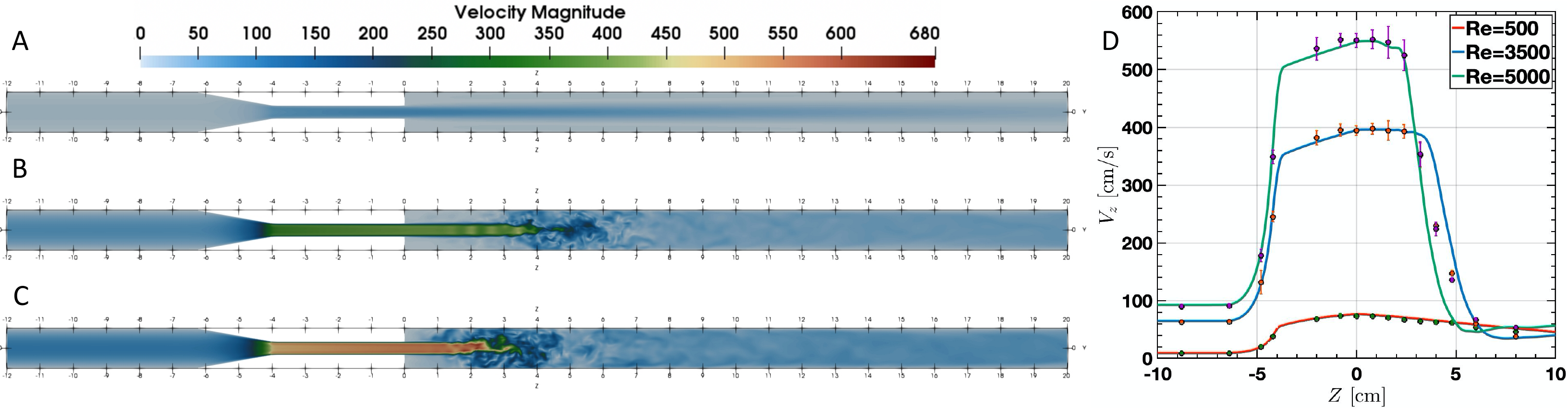}
\end{tabular}
\end{center}
\caption{CFD simulation results of an idealized medical device using the residual-based VMS formulation: Instantaneous velocity magnitude in cm/s for (A) $\textup{Re}=500$, (B) $\textup{Re}=3500$, and (C) $\textup{Re}=5000$. (D) Validation of these results against experimental data.} 
\label{fig:fda_benchmark_sanpshots}
\end{figure*}

\subsection*{Blood flow modeling}
Vascular flow patterns are strongly influenced by conduit geometries. Patient-specific hemodynamic modeling, which has opened the door to virtual treatment planning on a patient-by-patient basis, requires accurate geometric domains to be constructed from clinical image data, which are commonly acquired from computed tomography angiography or magnetic resonance imaging. Until recently, model construction has been a manual process requiring extensive user intervention in vessel centerline identification on a set of two-dimensional (2D) image slices, 2D image segmentation, and lofting of the segmentations into a three-dimensional (3D) anatomic bounded volume. Recent advances in machine learning show promise in accelerating the model building workflow \cite{Maher2019}. 

Upon construction of the computational domain, CFD, which has found extensive application in weather forecasting, visual effects in digital media, and aircraft and automobile design, can be used to simulate blood flow. The mathematical foundation of CFD is the Navier-Stokes equations, a set of nonlinear partial differential equations (PDEs) describing the conservation laws for viscous fluids. We note that while blood is in fact a suspension of blood cells in plasma that exhibits shear-thinning behavior, the Newtonian fluid model is widely accepted to be a good approximation in the large vessels. Referring back to the unified governing equations \eqref{eq:isothermal_governing_eqn_mass}-\eqref{eq:isothermal_governing_eqn_momentum}, incompressibility suggests a constant $\rho$ and $\beta_{\theta} = 0$, and the Newtonian fluid model suggests $\bm \sigma_{\mathrm{dev}} := 2\mu \bm \varepsilon_{\mathrm{dev}}$, in which $\mu$ is the dynamic viscosity, and $\bm \varepsilon_{\mathrm{dev}}$ is the deviatoric part of the rate-of-strain tensor. Furthermore, the mesh need not be moved for stationary fluid domains. As a result, the ALE coordinates $\bm \chi$ reduce to the Eulerian coordinates $\bm x$, and $\hat{\bm v} = \bm 0$. From a theoretical perspective, the nonlinear term in equation \eqref{eq:isothermal_governing_eqn_momentum} presents challenges associated with the existence and regularity of the solution of the Navier-Stokes equations, which remains among the Clay Mathematics Institute's unsolved Millennium Prize Problems \cite{fefferman2006}. A common approach has thus been to approximate the Navier-Stokes equations via high-fidelity numerical methods, such as finite element, finite volume, or finite difference methods, and extract physically meaningful information for analysis and prediction. Model reduction techniques, such as the reduced basis method for solving complex parameterized PDEs in low-dimensional spaces, have also been developed to reduce the computational cost.

The development of CFD techniques has been a major research thrust in applied mathematics and fluid mechanics in the past several decades. This subject is rich and merits a review article on its own; here we restrict comments to finite element formulations suitable for complex geometries and coupled problems. In recent years, the variational multiscale (VMS) formulation has received growing attention as a powerful technique for simulating 3D fluid problems. The VMS formulation is a priori consistent on all scales and can thus robustly achieve higher-order accuracy for both advection- and diffusion-dominated flows. In addition to its mathematical properties, numerical studies indicate that VMS properly captures statistical quantities relevant to flow physics, such as the energy spectrum and decay of kinetic energy in isotropic and wall-bounded turbulent flows \cite{hughes2018}. Like the classical stabilized methods, VMS further provides a mechanism to circumvent the Ladyzhenskaya-Babu{\v s}ka-Brezzi (LBB) condition associated with the divergence-free constraint, thus enabling arbitrary element types to be used for modeling incompressible flows. In contrast to \textit{LBB-stable} formulations \cite{Boffi2013} which require specific element types for velocity and pressure and is often implementationally inconvenient, these \textit{stabilized} formulations conveniently enable equal-order interpolation, largely simplifying both automatic mesh generation and finite element implementation from a practical standpoint. Indeed, the VMS formulation has been used in patient-specific blood flow simulations since the late 1990s. Figure \ref{fig:fda_benchmark_sanpshots} shows validation of simulated laminar, transitional, and fully turbulent flows in an idealized medical device against experimental data, demonstrating applicability of the VMS formulation for various flow regimes.

Over the years, the rise in high-performance computing and high-resolution imaging has been paralleled by significant advances in mathematical modeling techniques for flow simulations, and several open-source software packages are readily available. One representative example is the SimVascular project. In clinical practice, CFD has already been translated into decision-making in realms including diagnosis, risk stratification, and treatment planning. As a representative example, HeartFlow, Inc.  has achieved notable commercial success with its fractional flow reserve-CT technology for evaluating coronary lesions, which received clearance for clinical use from the US Food and Drug Administration (FDA) in 2019. By simulating the normalized pressure gradient through a coronary artery blockage, HeartFlow has eliminated the need for invasive coronary angiograms in 61\% of patient cases.

\begin{figure*}
	\begin{center}
	\begin{tabular}{c}
\includegraphics[angle=0, trim=0 100 0 0, clip=true, scale = 0.33]{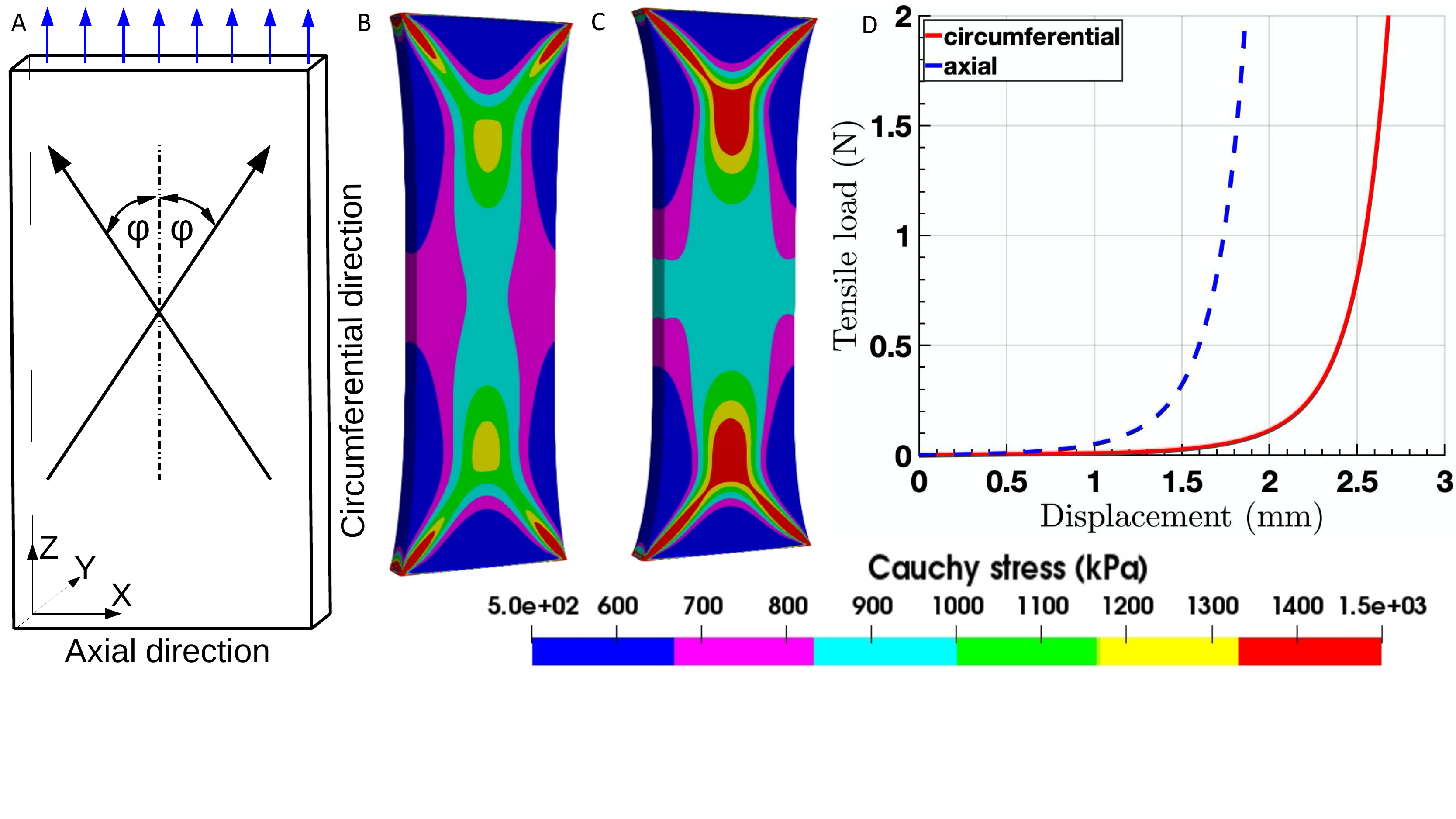}
\end{tabular}
\end{center}
\caption{Numerical tensile testing of a 3D collagen-reinforced vascular tissue model in the axial and circumferential directions. (A) Problem setting. (B) Resulting Cauchy stress distribution in an axially-tested specimen. (C) Resulting Cauchy stress distribution in a circumferentially-tested specimen. (D) Load-displacement curves demonstrating exponential growth in the stiffness provided by collagen fibers.} 
\label{fig:anisotropic}
\end{figure*}

\subsection*{Arterial wall modeling}
In addition to blood flow modeling, arterial wall modeling is equally critical to understanding the initiation and progression of cardiovascular diseases. Unlike engineering materials, biological tissues are highly deformable and behave as incompressible anisotropic visco-hyperelastic materials under physiological loading conditions. Moreover, biological tissues have been found to chronically adapt to the biomechanical environment through G\&R. From a modeling perspective, two distinct phenomena must be captured---the \textit{passive} nonlinear viscoelasticity and the \textit{active} G\&R of the soft tissue.

\begin{figure*}
	\begin{center}
	\begin{tabular}{c}
\includegraphics[angle=0, trim=230 420 150 420, clip=true, scale = 0.3]{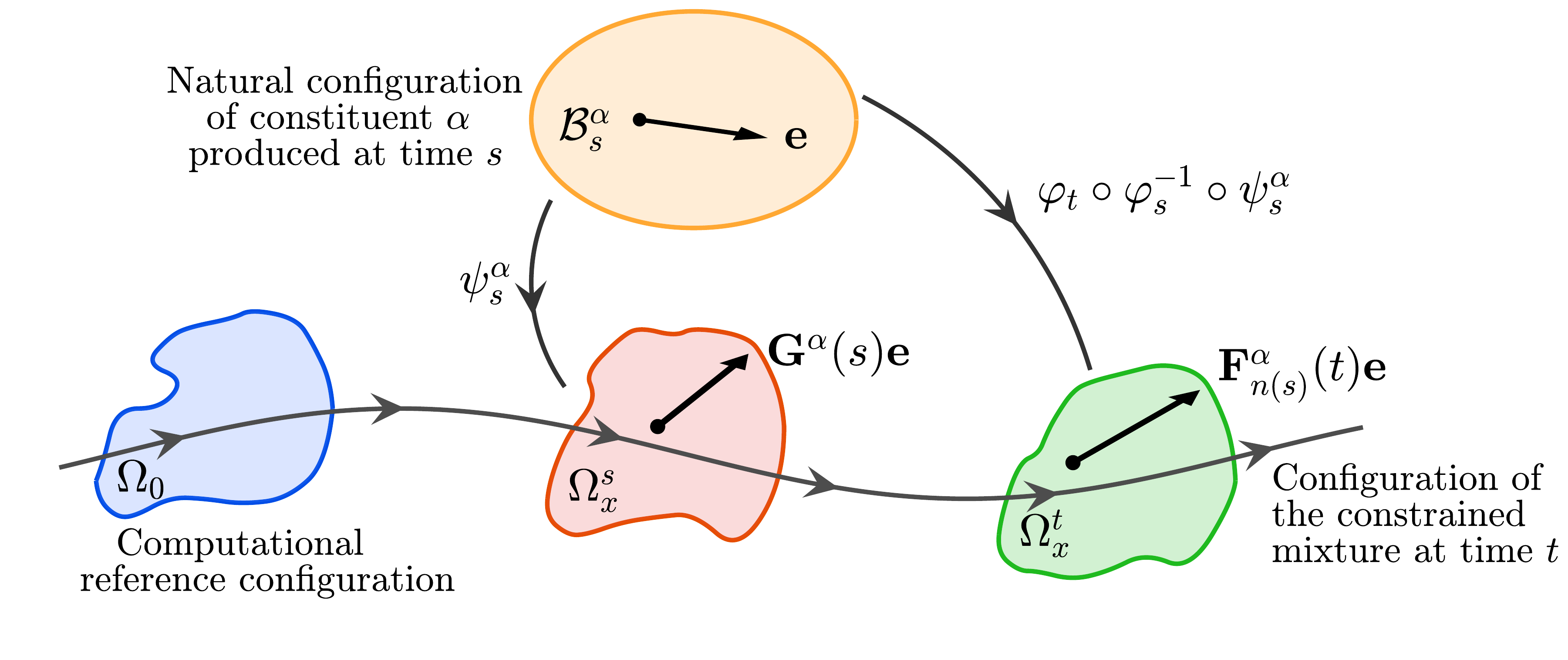}
\end{tabular}
\end{center}
\caption{Illustration of the various constrained mixture configurations. A vector $\bm e$ on $\mathcal B^{\alpha}_s$ is mapped to $\bm G^{\alpha}(s)\bm e$ on $\Omega_{\bm x}^{s}$ and $\bm F^{\alpha}_{s}(t)\bm e$ on $\Omega^{t}_{\bm x}$. } 
\label{fig:gnr_maps}
\end{figure*}

A Lagrangian framework is commonly adopted for the solid problem, as it enables accurate prescription of boundary conditions. Again referring back to the unified governing equations \eqref{eq:isothermal_governing_eqn_mass}-\eqref{eq:isothermal_governing_eqn_momentum}, the ALE coordinates $\bm \chi$ thus reduce to the Lagrangian coordinates $\bm X$, and $\hat{\bm v} = \bm v$. Furthermore, the displacement field $\bm u$ can conveniently be obtained by the kinematic relation $d\bm u/dt = \bm v$. Finally, the constitutive relations for $\rho$ and $\bm \sigma_{\mathrm{dev}}$ are given in terms of the Gibbs free energy $G$ as
\begin{align*}
& \rho = \left( \frac{\partial G}{\partial p} \right)^{-1}, \\
& \bm \sigma_{\mathrm{dev}} = J^{-1} \bm F \left(  2 \frac{\partial G}{\partial \bm C} \right) \bm F^T = J^{-1} \frac{\partial G}{\partial \bm F} \bm F^T,
\end{align*}
in which the deformation gradient $\bm F$, the Jacobian determinant $J$, and the right Cauchy-Green tensor $\bm C$ are given by
\begin{align*}
\bm F := \partial \bm u / \partial \bm X + \bm I, \quad J := \det\left( \bm F \right), \quad \bm C := \bm F^T \bm F.
\end{align*}
Computational modeling of incompressible nonlinear elasticity in complex geometries has long been a challenge. As mentioned above, the unified framework enables extension of the VMS formulation to the solid problem, thereby circumventing the LBB condition and enabling equal-order interpolation for incompressible finite elasticity. We do, however, note that LBB-stable formulations are equally attractive, as their energy stability can be demonstrated a priori in the following form,
\begin{align*}
& \frac{d}{dt} \int_{\Omega_{\bm X}} \frac12 \rho_0 \|\bm V_h\|^2 + \rho_0 G_{\mathrm{ich}}(\tilde{\bm C}_h) d\Omega_{\bm X} \\
& = \int_{\Omega_{\bm X}} \rho_0 \bm V_h \cdot \bm B d\Omega_{\bm X} + \int_{\Gamma_{\bm X}} \bm V_h \cdot \bm H d\Gamma_{\bm X},
\end{align*}
in which $\bm V_h$ and $\tilde{\bm C}_h$ are discrete approximations to $\bm V$ and $\tilde{\bm C} := J^{-2/3}\bm C$; $\bm V$, $\bm B$, and $\bm H$ are the velocity, body force, and surface traction in the Lagrangian configuration; and $G_{\mathrm{ich}} $ is the isochoric part of the Gibbs free energy. This energy stability estimate serves as an ``insurance plan" for the numerical scheme, guaranteeing reliability of the numerical results.

In addition to incompressibility, soft tissues also exhibit directionally dependent, or anisotropic, material properties as a result of collagen fiber orientations. The directional effect of a reinforcing fiber can be characterized by a structural tensor, and we can elegantly represent the energy potential $G$ using invariants of the strain tensor and structural tensors. Figure \ref{fig:anisotropic} illustrates the resulting Cauchy stress distributions of a 3D collagen-reinforced vascular tissue model subjected to numerical tensile testing in the circumferential and axial directions. Recently, we have additionally considered the inelastic response of soft tissues in our modeling of the \textit{passive} mechanical behavior. 

\begin{figure*}
	\begin{center}
	\begin{tabular}{c}
\includegraphics[angle=0, trim=50 20 50 20, clip=true, scale = 0.36]{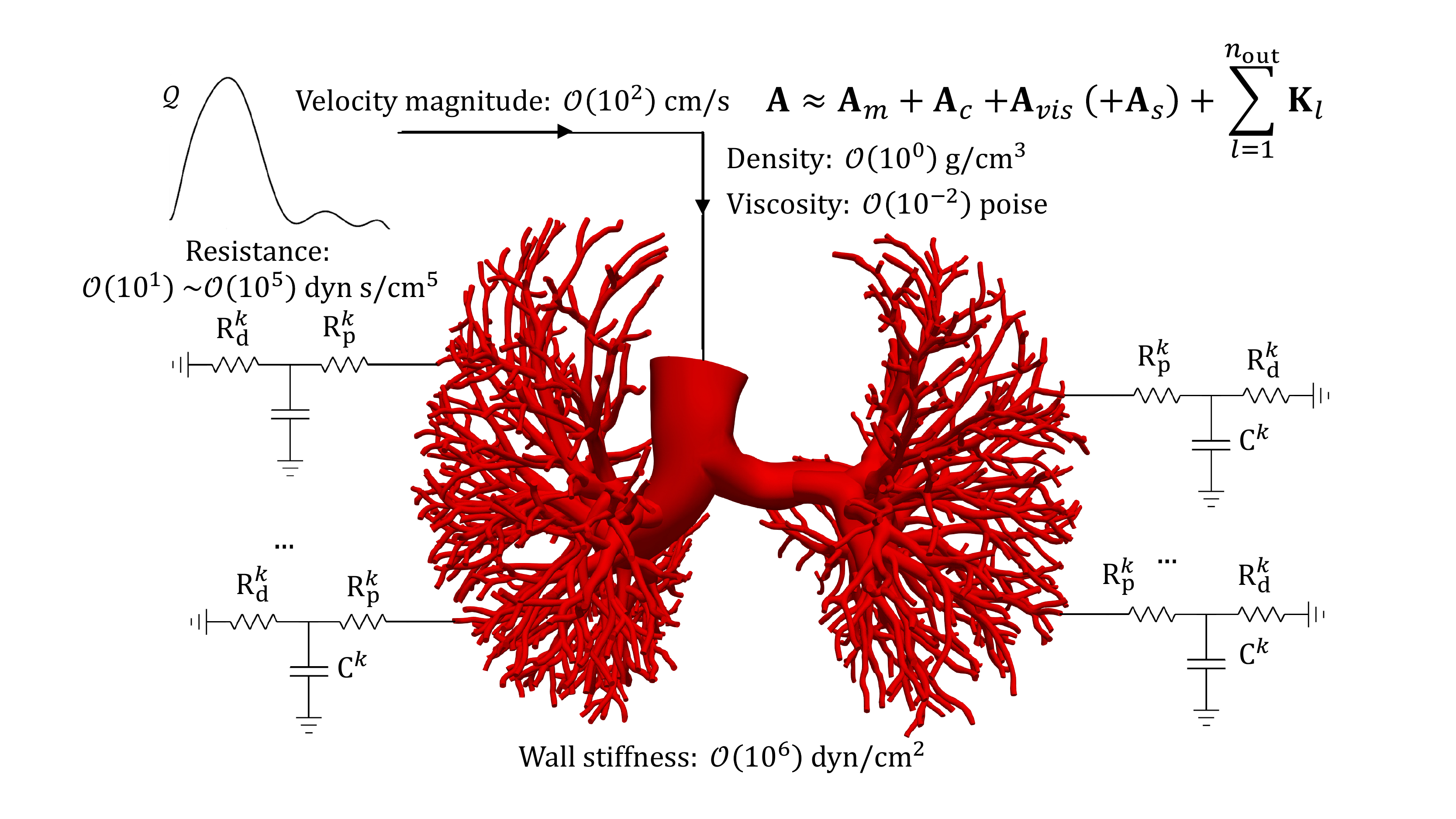}
\end{tabular}
\end{center}
\caption{Illustration of the structure of sub-matrix $\boldsymbol{\mathrm A}$ in cardiovascular FSI problems. The sub-matrix involves contributions from the transient term ($\boldsymbol{\mathrm A}_m$), convection term ($\boldsymbol{\mathrm A}_c$), viscous term ($\boldsymbol{\mathrm A}_{vis}$), wall stiffness ($\boldsymbol{\mathrm A}_s$), and boundary traction terms ($\boldsymbol{\mathrm K}_l$). These terms are scaled by physical parameters that span several orders of magnitude, reported here in the centimeter-gram-second units.} 
\label{fig:hemodynamics-algebra}
\end{figure*}

% contrained mixture model
The \textit{active} G\&R can be modeled by the constrained mixture theory, which was proposed to mathematically describe the evolution of arterial wall composition, morphology, and thus stiffness in response to sustained alterations in hemodynamic loading \cite{Humphrey2002}. The theory conceptualizes the arterial wall as a \textit{mixture} of structurally significant constituents, such as collagen, elastin, and smooth muscle cells. These constituents continually turnover at different production and removal rates, engendering the following set of evolution equations, in which the superscript $\alpha$ indexes the constituents,
\begin{align*}
\rho^{\alpha}(t) = \rho^{\alpha}(0) Q^{\alpha}(t) + \int_0^t \mathfrak m^{\alpha}(s) q^{\alpha}(s, t) ds.
\end{align*}
Here, $\rho^{\alpha}$ is the constituent's density, $Q^{\alpha}(t)$ is the fraction of the constituent present at time $t=0$ that survives to time $t$, $\mathfrak m^{\alpha}(s)$ is the production rate at time $s \in [0, t]$, and $q^{\alpha}(s, t)$ is the fraction of the constituent produced at time $s$ that survives to time $t$. The overall density of the tissue can then be defined as $\rho = \sum_{\alpha} \rho^{\alpha}$, and we can naturally define mass fractions as $\phi^{\alpha} := \rho^{\alpha}/\rho$. These evolution equations can be considered as an additional set of constitutive laws that yield source and sink terms for the mass balance equation \eqref{eq:isothermal_governing_eqn_mass}. To account for each constituent's contribution to the overall mechanical behavior, the energy potential is regarded as a mass-weighted average of the constituent potentials, $G = \sum_{\alpha} \phi^{\alpha} G^{\alpha}$.

Furthermore, the natural (or stress-free) configuration of each constituent is also allowed to evolve separately, suggesting that constituents produced at different times with different natural configurations can coexist. Newly produced constituents are deposited into the extant tissue at some prestretch, a process that can mathematically be described with a constituent-specific mapping $\psi^{\alpha}_s$ from the natural configuration $\mathcal B^{\alpha}_s$ to the tissue body $\Omega^s_{\bm x}$ at time $s$. We denote its corresponding tangent map as $\bm G^{\alpha}(s)$. The evolution of the tissue body is described with a separate mapping $\varphi_t$ from a material point $\bm X$ at time $0$ to a point $\bm x \in \Omega_{\bm x}^t$ at time $t$, such that all constituents are kinematically \textit{constrained} to deform together without any relative motion among the constituents. We denote the tangent map of $\varphi_t$ as $\bm F(t)$, also known as the deformation gradient. In practice, computing the tissue body's elastic stress requires a measure of the change in geometry from $\mathcal B^{\alpha}_s$ at time $s$ to $\Omega_{\bm x}^t$ at a later time $t$. This is achieved via a composition of the mappings, such that a point $\bm y \in \mathcal B^{\alpha}_s$ is mapped to a point $\bm x = \varphi_t \circ \varphi^{-1}_s \circ \psi^{\alpha}_s(\bm y) \in \Omega^t_{\bm x}$. We can finally represent the tangent map from $\mathcal B^{\alpha}_s$ to $\Omega^{t}_{\bm x}$ as 
\begin{align*}
\bm F^{\alpha}_{s}(t) = \bm F(t) \bm F^{-1}(s) \bm G^{\alpha}(s).
\end{align*}
Figure \ref{fig:gnr_maps} illustrates the kinematics of the constrained tissue mixture. To close the G\&R system, we define the deviatoric component of the Cauchy stress needed in the momentum balance equation \eqref{eq:isothermal_governing_eqn_momentum} as
\begin{align*}
\bm \sigma_{\mathrm{dev}}(t) = J^{-1}(t) \sum_{\alpha}  \frac{\partial \left(\phi^{\alpha} G^{\alpha}\right)}{\partial \bm F(t)} \bm F^T(t),
\end{align*}
in which the energy $\phi^{\alpha} G^{\alpha}$ is expressed through a hereditary integral,
\begin{align*}
\left( \phi^{\alpha} G^{\alpha} \right)(t) =& \frac{\rho^{\alpha}(0)}{\rho(t)} Q^{\alpha}(t) G^{\alpha}(\bm F^{\alpha}_0(t)) \\
&+ \int_0^t \frac{\mathfrak{m}^{\alpha}(s)}{\rho(t)} q^{\alpha}(s, t) G^{\alpha}(\bm F^{\alpha}_s(t)) ds.
\end{align*}
Studies of the constrained mixture theory demonstrate exciting potential for predicting the structural and morphological evolution observed in cardiovascular diseases, such as abdominal aortic aneurysms and myocardial hypertrophy. However, major challenges still exist for extension of this theory to 3D applications, as tracking the full evolutionary history may quickly become computationally intractable.

\subsection*{Linear solver}
Spatiotemporal discretization of nonlinear PDEs and the Newton-Raphson method eventually boil down to repeatedly solving a linear system with millions, or even billions, degrees of freedom. In computational science and engineering, solving linear systems often comprises the most time-consuming portion of the analysis and thus requires careful design. The local nature of discrete differential operators yields linear systems of sparse matrices that can be compactly stored in special representations, such as the compressed sparse row format. When used in conjunction with preconditioners, iterative methods such as Krylov subspace methods present the most effective solution procedure for sparse matrix problems on supercomputers. Importantly, the design of a preconditioner also dictates the overall efficiency, robustness, and scalability. Consider a fully implicit scheme for the PDEs \eqref{eq:isothermal_governing_eqn_mass}-\eqref{eq:isothermal_governing_eqn_momentum}, which yields a system of nonlinear algebraic equations that can be solved with the consistent Newton-Raphson method. The resulting linear system is associated with the following matrix $\boldsymbol{\mathcal A}$ of a $2 \times 2$ block structure, 
\begin{align*}
\boldsymbol{\mathcal A} =
\begin{bmatrix}
\boldsymbol{\mathrm A} & \boldsymbol{\mathrm B} \\
\boldsymbol{\mathrm C} & \boldsymbol{\mathrm D}
\end{bmatrix}.
\end{align*}
The sub-matrices $\boldsymbol{\mathrm A}$, $\boldsymbol{\mathrm B}$, and $\boldsymbol{\mathrm C}$ respectively represent the discrete convection-diffusion-reaction operator, gradient operator, and divergence operator, each with additional numerical modeling terms from the VMS formulation. The sub-matrix $\boldsymbol{\mathrm D}$ contains a mass matrix scaled by the isothermal compressibility coefficient and additional VMS modeling terms. Interested readers may refer to \cite{Liu2019} for explicit formulas of the sub-matrices. Block factorization may be performed on $\boldsymbol{\mathcal A}$ to yield matrices of lower triangular, diagonal, and upper triangular structure,
\begin{align*}
\boldsymbol{\mathcal A} = \boldsymbol{\mathcal L} \boldsymbol{\mathcal D} \boldsymbol{\mathcal U} =
\begin{bmatrix}
\boldsymbol{\mathrm I} & \boldsymbol{\mathrm O} \\
\boldsymbol{\mathrm C} \boldsymbol{\mathrm A}^{-1} & \boldsymbol{\mathrm I}
\end{bmatrix}
\begin{bmatrix}
\boldsymbol{\mathrm A} & \boldsymbol{\mathrm O} \\
\boldsymbol{\mathrm O} & \boldsymbol{\mathrm S}
\end{bmatrix}
\begin{bmatrix}
\boldsymbol{\mathrm I} & \boldsymbol{\mathrm A}^{-1} \boldsymbol{\mathrm B} \\
\boldsymbol{\mathrm O} & \boldsymbol{\mathrm I}
\end{bmatrix},
\end{align*}
in which $\boldsymbol{\mathrm S} := \boldsymbol{\mathrm D} - \boldsymbol{\mathrm C} \boldsymbol{\mathrm A}^{-1} \boldsymbol{\mathrm B}$ is the Schur complement. The design of a preconditioner for $\boldsymbol{\mathcal A}$ therefore reduces to solving smaller systems associated with $\boldsymbol{\mathrm A}$ and $\boldsymbol{\mathrm S}$. This design concept is closely related to the Chorin-Teman projection method and can be considered an algebraic procedure wrapping the projection method within an iterative solver. Whereas $\boldsymbol{\mathrm S}$ is as an algebraic manifestation of the pressure Poisson equation, $\boldsymbol{\mathrm A}$ contains several modeling contributions, as illustrated in Figure \ref{fig:hemodynamics-algebra} for cardiovascular FSI problems. These contributions in $\boldsymbol{\mathrm A}$ contain physical parameters that span several orders of magnitude, further complicating the definition of $\boldsymbol{\mathrm S}$, which involves the inverse of $\boldsymbol{\mathrm A}$ and is thus dense. The \textit{matrix-free} technique and \textit{multigrid} method represent two effective strategies that can be adopted. The matrix-free technique takes advantage of the fact that iterative methods do not require the algebraic definition of a matrix and instead only require the matrix's action on a vector. The non-sparse contributions therefore do not need to be explicitly constructed. As an example, we outline the matrix-free definition of $\boldsymbol{\mathrm S}$ in Algorithm \ref{algorithm:matrix_free_mat_vec_for_S}.
\begin{algorithm}[H]
\caption{The matrix-free algorithm for the multiplication of $\boldsymbol{\mathrm S}$ with a vector $\bm x$.}
\label{algorithm:matrix_free_mat_vec_for_S}
\begin{algorithmic}[1]
\State \texttt{Compute the matrix-vector multiplication} 

$\hat{\bm x} \gets \boldsymbol{\mathrm D} \bm x$.

\State \texttt{Compute the matrix-vector multiplication} 

$\bar{\bm x} \gets \boldsymbol{\mathrm B} \bm x$.

\State \texttt{Solve for $\tilde{\bm x}$ from the linear system} 

$\boldsymbol{\mathrm A} \tilde{\bm x} = \bar{\bm x}$.

\State \texttt{Compute the matrix-vector multiplication} 

$\bar{\bm x} \gets \boldsymbol{\mathrm C} \tilde{\bm x}$.

\State \Return $\hat{\bm x} - \bar{\bm x}$.
\end{algorithmic}
\end{algorithm}
\noindent The matrix-free method defines how the two sub-matrices can be solved using the Krylov subspace method. Nonetheless, to further accelerate convergence, preconditioners can be provided by constructing sparse approximations to the sub-matrices. Interestingly, both sub-matrices have significant contributions from elliptic operators. Their sparse approximations can thus be regarded as discrete elliptic operators and be addressed effectively by the multigrid method, which scales almost linearly with respect to the degrees of freedom for these types of matrices \cite{Elman2014}.

\begin{figure*}
	\begin{center}
	\begin{tabular}{c}
\includegraphics[angle=0, trim=0 260 0 0, clip=true, scale = 0.45]{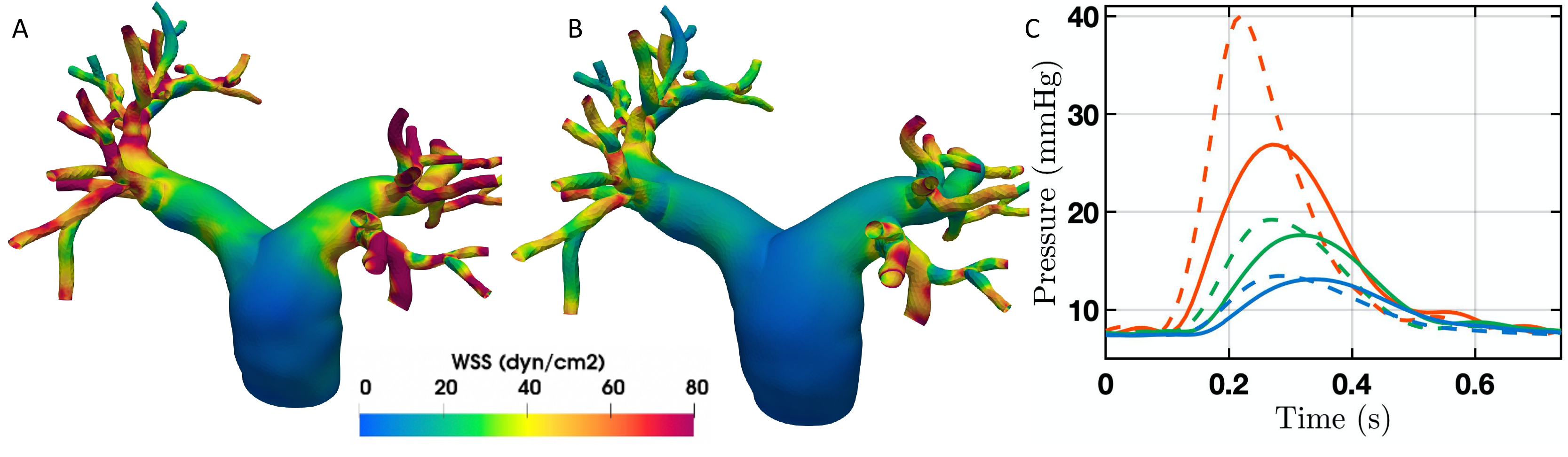}
\end{tabular}
\end{center}
\caption{Comparison of the WSS magnitude on a patient-specific pulmonary arterial model computed with (A) a rigid-wall simulation and (B) a FSI simulation. (C) Pressure over one cardiac cycle on the inlet (red) and two selected outlet surfaces (green and blue). Rigid-wall and FSI results are indicated by dashed and solid lines, respectively.} 
\label{fig:cfd-and-fsi}
\end{figure*}

In summary, the matrix $\boldsymbol{\mathcal A}$ is solved via a $\boldsymbol{\mathcal L} \boldsymbol{\mathcal D} \boldsymbol{\mathcal U}$ block decomposition that reduces the problem to solving the smaller block matrices $\boldsymbol{\mathrm A}$ and $\boldsymbol{\mathrm S}$, both of which can be solved iteratively without actual algebraic definitions. To accelerate convergence, sparse approximations of the two sub-matrices can be constructed to generate multigrid preconditioners without losing scalability on supercomputers.

\subsection*{Multiphysics modeling of FSI}
The coupling of biofluids and biosolids is of critical importance in cardiovascular modeling. The Navier-Stokes equations alone are insufficient for predicting the flow behavior, as such a model (sometimes  referred to as the rigid-wall model) neglects the deformation of the vessel wall, contractions of the heart, and motion of the heart valves, all of which give rise to a deforming fluid domain. Furthermore, the hemodynamic wall shear stress (WSS) and pressure are two important mechanical loads dictating the G\&R response of the vessel wall. As an illustrative example, Figure \ref{fig:cfd-and-fsi} compares the WSS magnitude and pressure computed from rigid-wall and FSI models. When neglecting the vessel wall compliance, hemodynamic quantities are consistently overpredicted.

\begin{figure*}
	\begin{center}
	\begin{tabular}{c}
\includegraphics[angle=0, trim=0 0 50 0, clip=true, scale = 0.32]{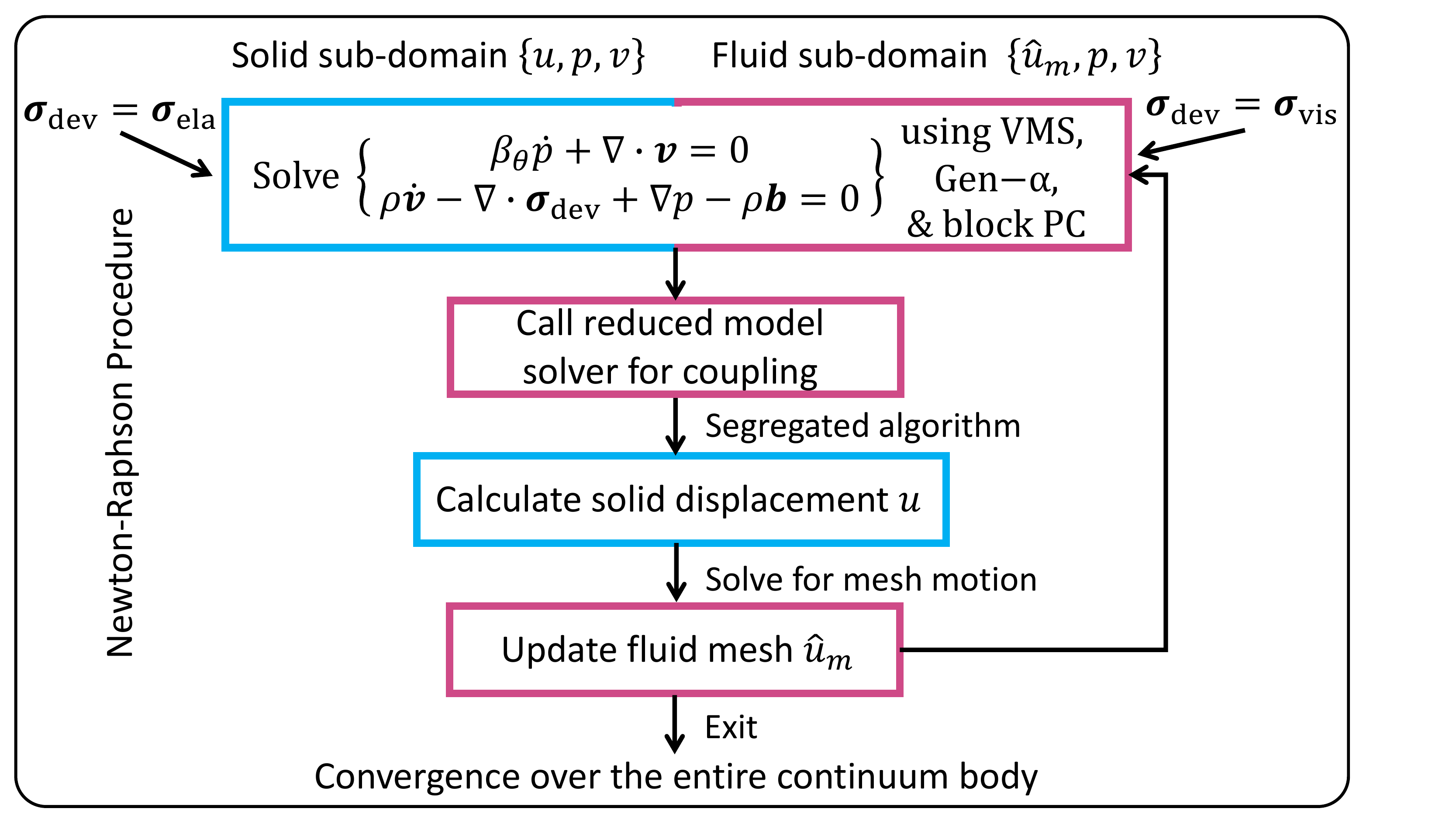}
\end{tabular}
\end{center}
\caption{Illustration of the Newton-Raphson procedure for a monolithically coupled FSI problem. The blue and magenta colors indicate procedures performed in the solid and fluid sub-domains, respectively. The solid displacement can be updated consistently using a segregated algorithm and subsequently used to determine the ALE mesh motion in the fluid sub-domain.} 
\label{fig:fsi_flowchart}
\end{figure*}

As discussed above, the unified continuum modeling framework enables us to describe both the fluid and solid sub-problems with the same set of governing equations, albeit with different forms for $\bm \sigma_{\mathrm{dev}}$, $\rho(p)$, and $\beta_{\theta}$. Given the deforming fluid domain in cardiovascular FSI problems, the fluid sub-problem requires an additional set of equations to describe the ALE mesh motion. Most commonly, the harmonic extension algorithm or the pseudo-linear-elasticity algorithm is used to determine $\hat{\bm v}$ in equation \eqref{eq:isothermal_governing_eqn_momentum}. A Lagrangian description is maintained for the solid sub-problem. Importantly, the unified framework enables monolithic coupling of the fluid and solid sub-problems with uniform spatiotemporal discretization via the VMS formulation and the generalized-$\alpha$ scheme. We note that the single resulting linear system for the FSI problem is precisely of the $2 \times 2$ block structure discussed above \cite{Liu2018}. As a result, the data structures and solution methods for the FSI problem can be constructed in a manner practically identical to that of CFD (Figure \ref{fig:fsi_flowchart}). To enable flexible coupling between dimensionally heterogeneous models (see the next section), a modular approach is adopted within each Newton-Raphson iteration to communicate information between the 3D and reduced models in a fashion similar to the Gauss-Seidel method \cite{Moghadam2013}.

\subsection*{Geometric multiscale modeling}
Given the computational expense associated with solving 3D problems, modeling the entire cardiovascular system with 3D models is intractable, even with modern-day computing facilities. In addition, the spatiotemporal resolution offered by 3D models is not always necessary for practical problems of interest. It is indeed possible to exploit the morphology of vessels to derive simplified models of reduced spatial resolution. These 1D or 0D reduced models can either be used as standalone models or as models coupled to a 3D model as boundary conditions.

\begin{figure}
	\begin{center}
	\begin{tabular}{c}
\includegraphics[angle=0, trim=20 86 410 80, clip=true, scale= 0.43]{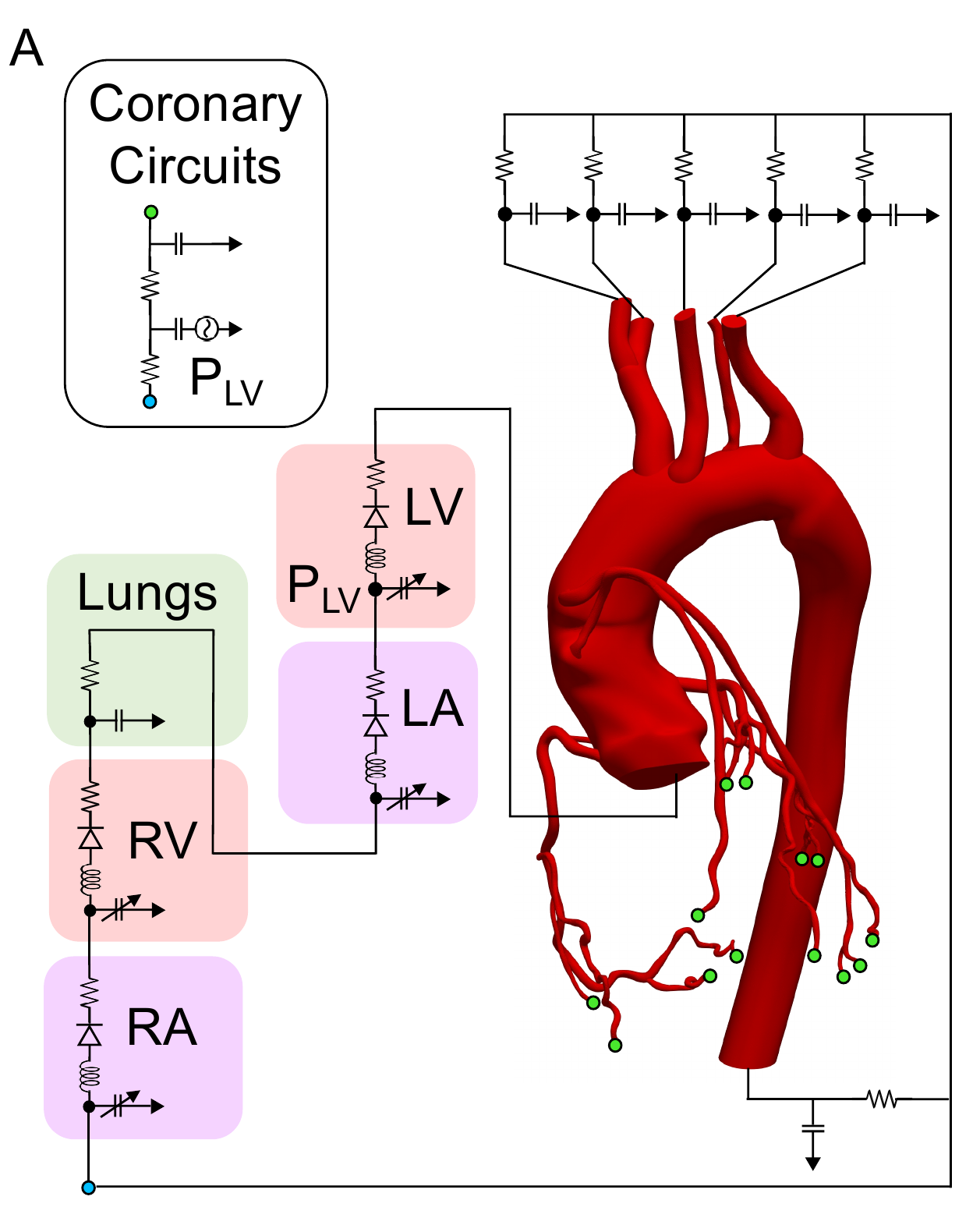}\\
\includegraphics[angle=0, trim=410 91 20 80, clip=true, scale= 0.43]{./figures/3D-1D-0D_coupling-v2.pdf}
\end{tabular}
\end{center}
\caption{Geometric multiscale modeling of the entire cardiovascular system as a closed-loop system. (A) 3D-0D coupling: the aorta and coronary arteries, which are generated from patient-specific image data, comprise the 3D domain; the peripheral vasculature and chambers of the heart (left and right atria and ventricles LA, RA, LV, RV) are modeled as 0D components serving as boundary conditions at the 3D inlet and outlets. (B) 3D-1D-0D coupling: The ascending aorta and coronary arteries are preserved in the 3D domain, while the remainder of the aorta and its daughter vessels are replaced with a 1D model generated from the vessel centerlines. } 
\label{fig:3D-1D-0D_coupling}
\end{figure}

In the 1D approach, which originated from Leonhard Euler's seminal work \cite{euler1862principia}, the 3D Navier-Stokes equations posed on a compliant axisymmetric tube are integrated over the cross-section, such that the spatial resolution is collapsed to a single axial dimension along the vessel,
\begin{align*}
& \frac{\partial A}{\partial t} + \frac{\partial Q}{\partial z} = 0, \\
& \rho \frac{\partial Q}{\partial t} + \rho \frac{\partial}{\partial z} \left( \alpha \frac{Q^2}{A} \right) + A\frac{\partial P}{\partial z} + K \frac{Q}{A} - \mu \frac{\partial^2 Q}{\partial z^2} = 0,
\end{align*}
where the three unknowns are the volumetric flow rate $Q$, the spatially-averaged pressure $P$, and the cross-sectional area $A$. Here, the parameters $\alpha$ and $K$ depend on the assumed velocity profile over the cross-section. For a parabolic profile, for example, the momentum flux correction coefficient $\alpha = 4/3$, and the friction parameter $K = 8\pi \mu$. A constitutive wall model describing the functional dependence of $P$ on $A$, such as the following simple algebraic relation, is required to close the system,
\begin{align*}
& P(z, t) = P_{\mathrm{ext}}(z, t) + \psi(A(t, z)), \\
& \psi(A) = \frac{Eh\sqrt{\pi}}{1 - \nu^2} \frac{\sqrt{A} - \sqrt{A_0}}{A_0},
\end{align*}
where $P_{\mathrm{ext}}$ is the external pressure, $A_0$ is the reference area when $P = P_{\mathrm{ext}}$, $E$ is the Young's modulus, $h$ is the thickness, and $\nu$ is the Poisson's ratio. Assuming $A > 0$, the above system is strictly hyperbolic with the following two distinct real eigenvalues, 
\begin{align*}
& \lambda_{1, 2} = \alpha\frac{Q}{A} \pm \sqrt{c^2 + \left(\frac{Q}{A}\right)^2\alpha (\alpha-1)}, \\ 
& c := \sqrt{\frac{A}{\rho}\frac{d \psi(A)}{d A}} = \sqrt{\frac{Eh\sqrt{\pi A}}{2\rho(1 - \nu^2)A_0}}.
\end{align*}
Since $c \gg \alpha Q/A$ in hemodynamic applications, the waves travel in two distinct directions. To properly capture the wave propagation phenomena, the Lax-Wendroff scheme or discontinuous Galerkin method can be used to solve this set of nonlinear hyperbolic equations.

The spatial resolution can be eliminated altogether by further integrating the 1D system over a segment of the vessel with length $l$ and assuming a small Reynolds number such that the nonlinear convective term can be neglected. With the above choice for $\psi(A)$, the resulting 0D model can be derived as follows,
\begin{align*}
& C\frac{d\mathcal{P}}{dt} + \mathcal{Q}_{\mathrm d} - \mathcal{Q}_{\mathrm p} = 0, \\
& L\frac{d\mathcal{Q}}{dt} + R\mathcal{Q} + \mathcal{P}_{\mathrm d} - \mathcal{P}_{\mathrm p} = 0,
\end{align*}
wherein
\begin{align*}
R :=\frac{K l}{A_0^2}, \quad C := \frac{2A_0\sqrt{A_0} (1 - \nu^2) l}{Eh\sqrt{\pi}}, \quad L := \frac{\rho l}{A_0},
\end{align*}
represent the viscous resistance, vessel compliance, and blood inertance, respectively. Here, the two unknowns are the longitudinally averaged flow $\mathcal Q$ and pressure $\mathcal{P}$, and the subscripts $\mathrm p$ and $\mathrm d$ denote the proximal and distal values. The above equations also arise in electrical circuits or hydraulic networks, and it is indeed popular to establish an electric-hydraulic analogy, in which the current is analogous to flow, and voltage to pressure. The 0D models can be constructed as arbitrary combinations of resistance, capacitance, and inductance elements in series and in parallel with proper matching conditions. Despite the lack of any spatial dependence and thus the failure to capture wave propagation phenomena, 0D models comprised of compartments representing distinct portions of the vasculature or chambers of the heart are often sufficient for approximating flow dynamics in the global circulation. Furthermore, the reduction of the 3D governing equations to ordinary differential-algebraic equations reduces the computational time by several orders of magnitude. As is true for many modeling techniques, there is a clear trade-off between accuracy and speed.

Regardless of the choice of model, boundary conditions must be properly assigned. In particular, boundary conditions for a 3D FSI model must be consistent with the wave propagation dynamics without introducing spurious reflections into the 3D domain. Reduced models of the downstream vasculature thus represent an extremely fitting choice. Geometric multiscale modeling, or the coupling of dimensionally heterogeneous models, further offers an efficient approach for modeling the cardiovascular system, in which the vast majority of computing resources are reserved for 3D FSI modeling of a localized region of interest. Figure \ref{fig:3D-1D-0D_coupling} illustrates both 3D-0D and 3D-1D-0D coupling. In the latter, a portion of the initial 3D domain is replaced with a 1D model generated from the vessel centerlines. We note that just as 3D FSI problems require appropriate transmission conditions on the fluid-solid interface, geometric multiscale modeling requires mathematically and physically sound transmission conditions. Interested readers may refer to \cite{quarteroni2019mathematical} for an elaboration on this topic.

\subsection*{Optimization and uncertainty quantification}
Given reliable solution techniques for the forward problem of a multiscale multiphysics system, one can further invoke the mathematical concept of optimization to improve designs for medical devices and surgical interventions (commonly referred to as shape optimization) or to identify modeling parameters. Conceptually, optimization seeks to minimize or maximize an objective function, given a set of input parameters subject to certain constraints. Examples of shape optimization include optimization of the radii, attachment angles, and attachment locations of shunts and grafts in single-ventricle palliation, such that maximized pulmonary flow and an unskewed hepatic flow distribution are achieved while preserving low pressures in the vena cava \cite{Yang2013}. On the other hand, optimization can also be used as a tool for data assimilation, which seeks to identify modeling parameters that best reproduce experimental or clinical data. For example, external tissue support parameters have previously been identified by minimizing the discrepancy between model and image contours of the vessel lumen, and G\&R modeling parameters have been identified to achieve homeostatic stress states experimentally observed in the vessel. In all cases, appropriate objective functions must be determined to measure performance, a process that requires deep understanding of the clinical disease or mechanobiological system of interest.

\begin{figure}
	\begin{center}
	\begin{tabular}{c}
\includegraphics[angle=0, trim=50 0 380 0, clip=true, scale= 0.33]{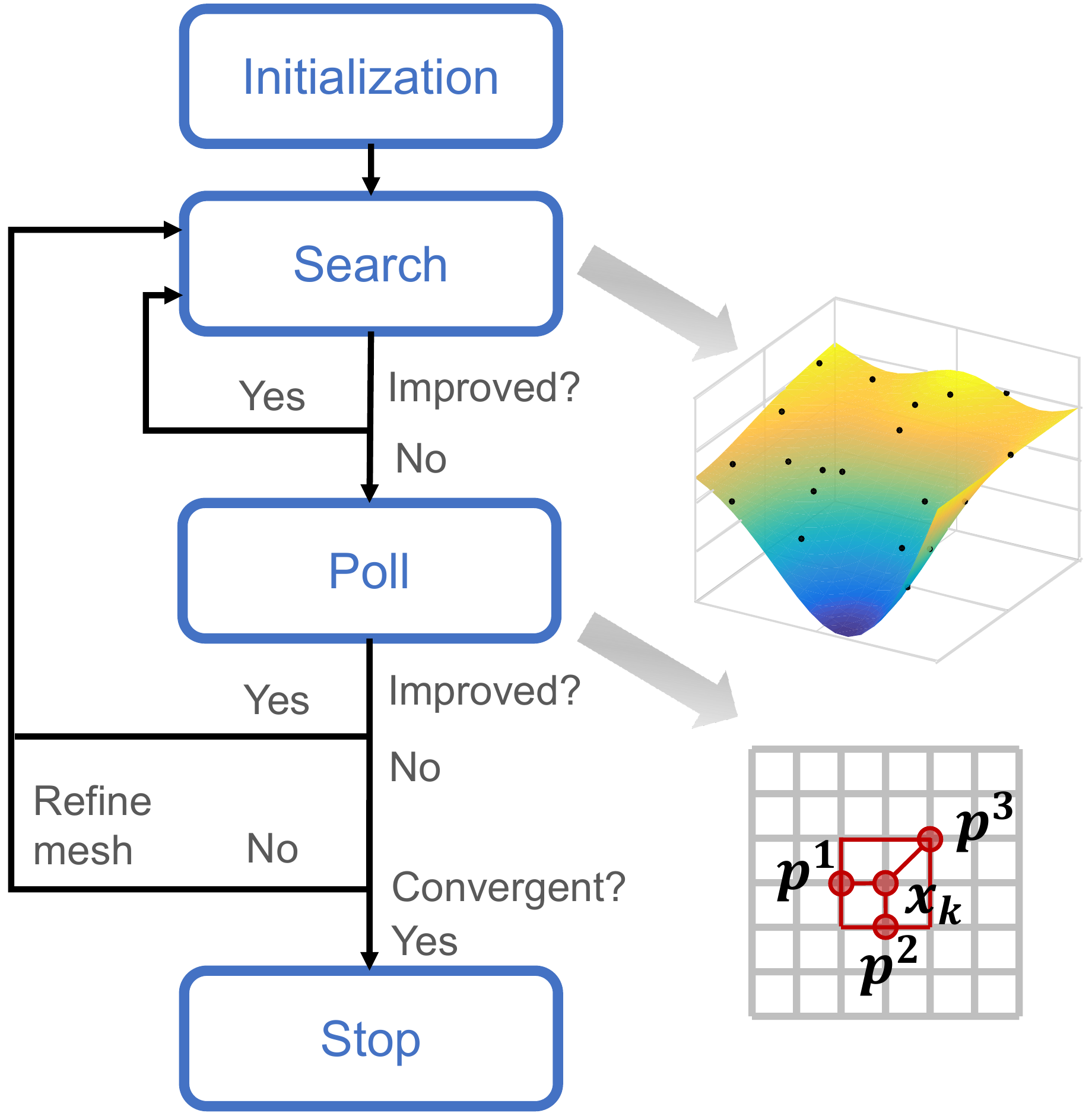}
\end{tabular}
\end{center}
\caption{Flowchart for shape optimization using the surrogate management framework.} 
\label{fig:SMF-flowchart}
\end{figure}

\begin{figure}
\begin{center}
\includegraphics[angle=0, trim=120 115 160 115, clip=true, scale= 0.32]{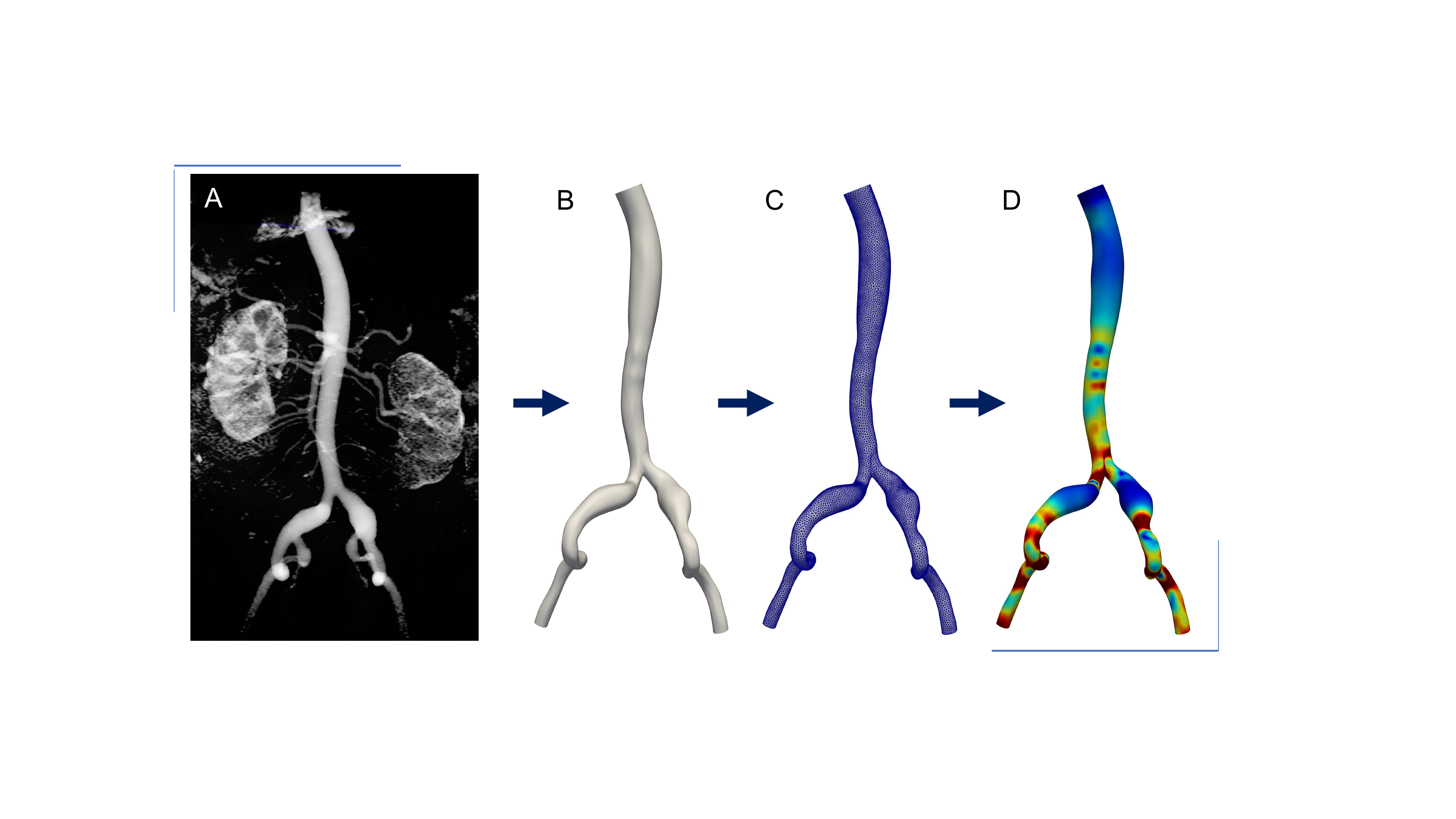}
\end{center}
\caption{The SimVascular pipeline (from left to right): medical image segmentation, model construction, mesh generation, and simulation.} 
\label{fig:sv-pipeline}
\end{figure}

Coupling cardiovascular simulations to optimization algorithms is particularly challenging, as each objective function evaluation requires a solution of the forward problem. Given the computational cost and complexity, the application of conventional gradient-based optimization methods, which require numerical determination of gradients via the finite difference or adjoint-state method, remains cost prohibitive. The surrogate management framework (SMF) has thus emerged as a black-box derivative-free method robust for solving expensive cardiovascular optimization problems. The SMF is comprised of two essential components. The SEARCH step uses a surrogate function, or an approximation of the objective function with reduced computational cost, to identify design points that are likely to improve the objective function; the POLL step evaluates points in a positive spanning set of directions around the current best point in the parameter space (Figure \ref{fig:SMF-flowchart}). It therefore combines the efficiency of methods based on response surfaces with the convergence properties of pattern search methods \cite{Marsden2014}. However, we note that to date, cardiovascular problems have been limited to idealized geometries, as automation of the entire modeling, meshing, and simulation pipeline remains challenging for complex geometries that are difficult to parameterize.

Given the large number of inputs to cardiovascular simulations, several sources of uncertainty exist. These pertain to clinical measurements, image-based geometries, material properties, and boundary conditions, just to name a few. As simulations are increasingly incorporated into the FDA approval process for medical devices and diagnostic tools, rigorously assessing the impact of uncertainty on simulation predictions must be considered a priority. Uncertainty quantification (UQ) is precisely the mathematical field that seeks to propagate these input uncertainties forward, such that simulation outputs are ultimately quantified with probability density functions and confidence intervals. Monte Carlo (MC) sampling, one of the first approaches proposed for UQ, is unbiased and flexible with arbitrary and potentially correlated inputs. Despite these appealing features, its slow convergence rate makes it cost prohibitive, especially considering the computational expense associated with 3D simulations. Recent work has extended MC to multilevel multifidelity MC, which makes use of different spatial resolutions (mesh size) and model fidelities (3D, 1D, 0D), to achieve reduced variance for a fixed computational cost \cite{Fleeter2020}. As an alternative to MC, polynomial chaos expansions describe smooth stochastic responses using interpolating polynomials that are mutually orthogonal with respect to the probability measure of the random inputs and can achieve up to exponential convergence rates with respect to the polynomial order. Significant research has also focused on adaptive methods appropriate for applications involving discontinuous stochastic responses. We conclude this section by emphasizing that optimization and UQ are two important mathematical tools that, when combined, enable identification of optimal designs and parameters that are robust to fluctuations in modeling choices and surgical implementation \cite{Sankaran2010}.

\subsection*{The SimVascular project}
\begin{figure*}
	\begin{center}
	\begin{tabular}{c}
\includegraphics[angle=0, trim=10 60 130 10, clip=true, scale= 0.34]{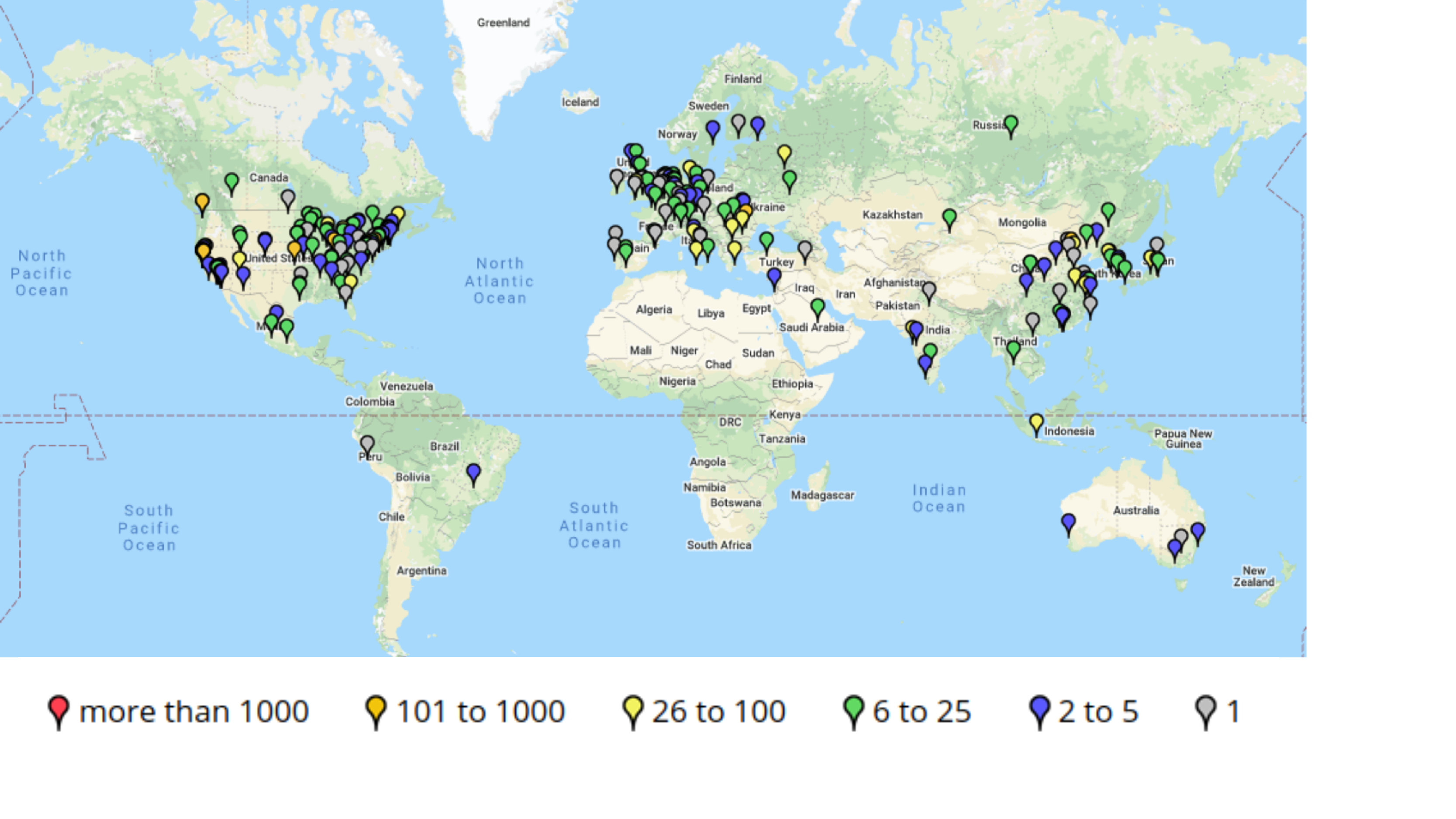}
\end{tabular}
\end{center}
\caption{Geographical distribution of SimVascular users.} 
\label{fig:sv-stat}
\end{figure*}

To date, SimVascular \cite{simvascular} remains the only fully open-source software offering a complete pipeline (Figure \ref{fig:sv-pipeline}) from medical image segmentation to 3D model construction, meshing, and patient-specific vascular FSI simulation with either ALE or a reduced linear membrane formulation \cite{Updegrove2017}. It was originally developed in Charles Taylor's lab at Stanford University and released in 2007. At the time, however, the integration of several licensed commercial components hindered new user adoption and prevented complete open source release. In 2013, the senior author of this article accordingly launched a joint revitalization effort with collaborators, aiming to integrate open-source alternatives and improve computational techniques for all stages of the pipeline. Nowadays, SimVascular has attracted over 4500 domestic and international users (Figure \ref{fig:sv-stat}). The development team continues to host conference workshops, Youtube tutorials, and a user forum to foster further advances in cardiovascular research. It has also proven to be an effective educational tool in graduate engineering courses, elucidating fundamental principles in CFD, FSI, and human physiology. As an active software project with regularly forthcoming enhancements, it represents the state of the art in cardiovascular simulation. Current research and development efforts span finite element method development for 3D G\&R, automated vessel path identification, model manipulation, optimization, UQ, and even virtual reality for visualizing simulation results in real-time. Importantly, the motivation underlying these efforts is the ultimate vision of virtual patient-specific treatment planning on a clinically relevant time frame, in which physicians can make informed decisions about the optimal intervention directly from the clinic.

\subsection*{Open challenges and future directions}
Having summarized the state of the art in mathematical modeling of the vascular system, we make use of this final section to discuss open challenges worthy of further investigation.

\paragraph{Multiphysics and multiscale modeling.}
Given the dynamic interplay among hemodynamics, vascular wall biomechanics, and cellular biochemical responses, there is a compelling need for mathematical models characterizing the impact of hemodynamics on mechanotransduction pathways and thus the wall composition and material properties. In addition to the deformation induced by mechanical loads, cell-mediated changes involved in G\&R must also be modeled. Such a mathematical view of all scales is critical for directly translating disease conditions into models with quantifiable parameters and outputs. For example, predictive modeling of thrombus formation requires the flow-mediated transport phenomena to be integrated with microscale coagulation kinetics. More refined multiphysics and multiscale mathematical models are indeed needed for clinical applications.

\paragraph{Geometric parameterization.}
Despite advances in optimization algorithms for cardiovascular simulations, practical optimization applications with complex geometries are still hindered by the lack of geometric parameterization and manipulation techniques compatible with existing computer-aided design (CAD) frameworks. Extending optimization studies beyond idealized geometries would require direct manipulation of meshes without introducing discontinuities or severe mesh distortion. Recent efforts in combining CAD with computer-aided engineering (CAE) demonstrate great promise for seamless integration and automation of geometric manipulation and physics-based simulations \cite{Hughes2010}.

\paragraph{Verification and validation.}
As mathematical models for multiphysics phenomena become increasingly sophisticated, software implementations correspondingly grow in complexity. A recent study assessing the variability among CFD results from several research groups found rather large discrepancies for a relatively simple clinical test case \cite{Sendstad2018}. The results from this multi-laboratory challenge signified that in the face of wide adoption of medical simulation technology, there is a pressing need for rigorous verification of numerical techniques and equally rigorous validation of mathematical models. A close collaboration among applied mathematicians, engineers, experimentalists, and clinicians is essential for establishing an accreditation system for mathematical and computational modeling of the vascular system.

\subsection*{Acknowledgments}
The authors acknowledge Melody Dong for the preparation of Figure \ref{fig:hemodynamics-algebra}. This work is supported by the Guangdong-Hong Kong-Macao Joint Laboratory for Data-Driven Fluid Mechanics and Engineering Applications under the award number 2020B1212030001, the National Institutes of Health grants R01HL139796 and R01EB018302, and the National Science Foundation grant 1663671.

\bibliography{nams}
\end{document}